\documentclass[conference]{IEEEtran}
\IEEEoverridecommandlockouts
\usepackage{cite}
\usepackage{amsmath,amssymb,amsfonts}
\usepackage{algorithmic}
\usepackage{graphicx}
\usepackage{textcomp}
\usepackage[table]{xcolor}  
\usepackage{multirow}
\usepackage{bm}
\usepackage{subfigure}
\usepackage{makecell}
\usepackage{url}
\usepackage{tcolorbox}
\usepackage{soul}
\usepackage{algorithm}
\usepackage{ulem}
\usepackage{hyperref}

\newcommand{\graycell}[1]{\colorbox[gray]{0.9}{\makebox[2em][r]{#1}}}
\newcommand{\whitecell}[1]{\colorbox{white}{\makebox[2em][r]{#1}}}

\def\BibTeX{{\rm B\kern-.05em{\sc i\kern-.025em b}\kern-.08em
    T\kern-.1667em\lower.7ex\hbox{E}\kern-.125emX}}
\begin{document}

\title{A Deep Dive into Retrieval-Augmented Generation for Code Completion: Experience on WeChat}


\makeatletter
\newcommand{\IEEEauthorbox}[4]{%
  \parbox{0.31\textwidth}{\centering
    {\@IEEEauthorblockNstyle #1}\\
    {\@IEEEauthorblockAstyle \textit{#2}\\#3\\#4}%
  }%
}
\makeatother

\author{%
\IEEEauthorbox{Zezhou Yang}{Tencent}{Guangzhou, China}{zezhouyang@tencent.com}
\hfill
\IEEEauthorbox{Ting Peng}{Tencent}{Guangzhou, China}{sakurapeng@tencent.com}
\hfill
\IEEEauthorbox{Cuiyun Gao$^{*}$}{The Chinese University of Hong Kong}{Hong Kong, China}{cuiyungao@outlook.com}
\\[1em]
\IEEEauthorbox{Chaozheng Wang}{The Chinese University of Hong Kong}{Hong Kong, China}{adf111178@gmail.com}
\hfill
\IEEEauthorbox{Hailiang Huang}{Tencent}{Guangzhou, China}{eraserhuang@tencent.com}
\hfill
\IEEEauthorbox{Yuetang Deng}{Tencent}{Guangzhou, China}{yuetangdeng@tencent.com}
\thanks{$^{*}$Cuiyun Gao is the corresponding author.}
\thanks{This research is supported by National Key R\&D Program of China (No. 2022YFB3103900), National Natural Science Foundation of China under project (No. 62472126), Natural Science Foundation of Guangdong Province (Project No. 2023A1515011959), Shenzhen-Hong Kong Jointly Funded Project (Category A, No. SGDX20230116091246007), and Shenzhen Basic Research (General Project No. JCYJ20220531095214031).}
}

\maketitle

\begin{abstract}
Code completion, a crucial task in software engineering that enhances developer productivity, has seen substantial improvements with the rapid advancement of large language models (LLMs). 
In recent years, retrieval-augmented generation (RAG) has emerged as a promising method to enhance the code completion capabilities of LLMs, which leverages relevant context from codebases without requiring model retraining. 
While existing studies have demonstrated the effectiveness of RAG on public repositories and benchmarks, the potential distribution shift between open-source and closed-source codebases presents unique challenges that remain unexplored.
To mitigate the gap, we conduct an empirical study to investigate the performance of widely-used RAG methods for code completion in the industrial-scale codebase of WeChat, one of the largest proprietary software systems. Specifically, we extensively explore two main types of RAG methods, namely identifier-based RAG and similarity-based RAG, across 26 open-source LLMs ranging from 0.5B to 671B parameters.
For a more comprehensive analysis, we employ different retrieval techniques for similarity-based RAG, including lexical and semantic retrieval. Based on 1,669 internal repositories, we achieve several key findings:
(1) both RAG methods demonstrate effectiveness in closed-source repositories, with similarity-based RAG showing superior performance,
(2) the effectiveness of similarity-based RAG improves with more advanced retrieval techniques, where BM25 (lexical retrieval) and GTE-Qwen (semantic retrieval) achieve superior performance,
and (3) the combination of lexical and semantic retrieval techniques yields optimal results, demonstrating complementary strengths.
Furthermore, we conduct a developer survey to validate the practical utility of RAG methods in real-world development environments.
\end{abstract}


\section{Introduction}
\label{sec-1}
Code completion, which automatically predicts and suggests code fragments based on the surrounding programming context, has evolved from simple token-level suggestions to generating entire code blocks \cite{liu2024graphcoder,wang2024teaching}. Studies have demonstrated that code completion tools substantially enhance developer productivity in real-world software development \cite{DBLP:conf/icse/IzadiKDOPD24,repobench}. Notably, 87\% of professional developers report significant improvements in their coding efficiency when utilizing code completion tools in industrial settings \cite{DBLP:conf/sigsoft/WangHGJ0HLD23}.
Recent advances in large language models (LLMs) have further transformed various software engineering tasks \cite{DBLP:conf/kbse/GaoWGWZL23,DBLP:conf/kbse/PengWWGL23,DBLP:conf/icse/Du0WWL0FS0L24,DBLP:conf/acl/Du0ZXJLS024,daneshvar2024exploring}, demonstrating unprecedented capabilities in code understanding and generation. 
These models have achieved particularly impressive performance in code completion tasks \cite{repofuse,repoformer,24-Cheng}.

To enhance LLMs' performance on domain-specific tasks, researchers have explored Retrieval-Augmented Generation (RAG), which augments model inference by retrieving and incorporating relevant context from the target codebase without requiring parameter updates \cite{gao2023retrieval,zhang2024rag}.
The emergence of RAG methods provides a promising approach to leverage the powerful abilities of LLMs for industrial software development \cite{ahmed2024studying}. It not only preserves the privacy of proprietary code but also enables models to adapt to specific coding styles.
While RAG for code completion has shown promising results on public repositories and benchmarks \cite{reacc,FT2Ra}, the characteristics of closed-source codebases present unique challenges. Closed-source repositories often contain proprietary code patterns, custom frameworks, and domain-specific implementations that differ from open-source codebases \cite{andersen2012commercial}. The inherent differences raise practical concerns about the applicability of RAG methods in proprietary settings. However, a systematic study of RAG for code completion in closed-source repositories remains unexplored, leaving practitioners without clear guidance for leveraging RAG in proprietary development environments.

To mitigate the gap, we conduct a comprehensive investigation of RAG-based code completion in industrial settings, using the proprietary codebase from the WeChat Group in Tencent. 
As one of the largest social platforms with over 1 billion monthly active users \cite{activemonthly}, WeChat maintains a sophisticated codebase with
internal well-defined development practices and complex business logic, providing a compelling real-world environment for evaluating RAG methods at an industrial scale.
To enable systematic evaluation of RAG methods, we conduct two essential preparatory steps. 
First, we construct a carefully curated evaluation benchmark comprising 100 examples across seven domains, with manually annotated context and comments that reflect real-world code completion scenarios. Second, we collect 1,669 internal repositories from WeChat's development ecosystem as the source of the retrieval corpus. Following this, 
we propose a data preprocessing algorithm that extracts multiple pieces of context information to construct a fine-grained retrieval corpus \cite{arusoaie2017comparison,hatledal2019language}.

Our experiments evaluate 26 open-source LLMs ranging from 0.5B to 671B parameters, thoroughly investigating the capabilities of RAG methods in closed-source scenarios. Specifically, we address three key research questions:

\textbf{RQ1: How do different RAG methods perform in closed-source code completion?} This research question compares two types of RAG paradigms: identifier-based RAG, which retrieves relevant definitions of identifiers to help LLMs understand their inner logic and usage; and similarity-based RAG, which retrieves similar code implementations using lexical (BM25) and semantic (CodeBERT, UniXcoder, CoCoSoDa, and GTE-Qwen) retrieval techniques.
Our experimental results demonstrate the effectiveness of both types of methods, with similarity-based RAG exhibiting superior performance across different model scales.

\textbf{RQ2: How do the retrieval techniques affect similarity-based RAG on code completion?} 
We investigate the impact of using different query formulations (incomplete code context and complete code snippets) across various retrieval techniques within similarity-based RAG. 
Our experimental results reveal that lexical retrieval consistently shows strong performance across different code completion models, while the effectiveness of semantic retrieval scales positively with model capacity. Moreover, most retrieval techniques perform better with complete code snippets, whereas GTE-Qwen demonstrates superior performance with incomplete code context.

\textbf{RQ3: What is the relationship between different types of retrieval techniques in similarity-based RAG?} To answer this research question, we first conduct a comparative analysis between lexical (BM25) and semantic retrieval techniques by examining their retrieved results. Despite achieving similar performance levels individually, we discover minimal overlap in their retrieved candidates, which suggests they capture fundamentally different aspects of code similarities.
Building upon this observation, we find that combining BM25 with GTE-Qwen yields optimal performance across most LLMs, demonstrating the value of hybrid approaches in RAG-based code completion.

\begin{figure}[h]
  \centering
  \includegraphics[width=0.85\linewidth]{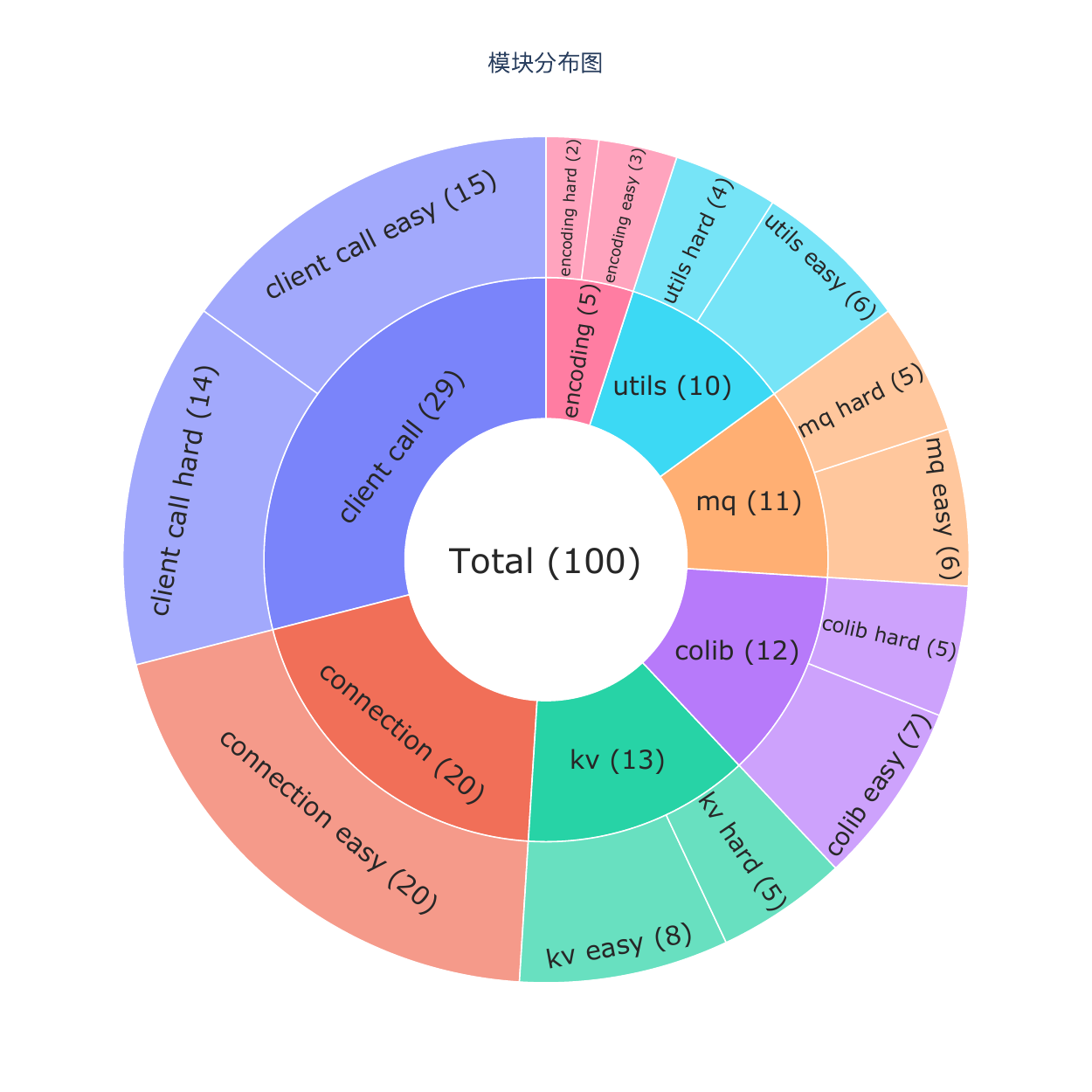}
  \caption{Statistics of our benchmark.}
  \label{fig:benchmark}
\end{figure}

The paper makes the following contributions:
\begin{enumerate}
    \item We conduct a systematic study of retrieval-augmented code completion on closed-source codebases, providing comprehensive empirical insights into the effectiveness of different RAG methods using 26 open-source LLMs.
    \item We propose a data preprocessing algorithm for constructing a fine-grained retrieval corpus from large-scale codebases, addressing the challenge of context extraction in RAG for code completion.
    \item Our experiment results reveal the complementary nature of lexical and semantic retrieval techniques, demonstrating that their combination can further enhance RAG-based code completion performance.
    \item We validate our empirical findings through a developer survey, confirming that the observed performance improvements align with developers' practical experiences in real-world scenarios.
\end{enumerate}

\section{Retrieval-Augmented Code Completion}
\label{sec-2}
\subsection{Benchmark Construction}
To comprehensively evaluate RAG's performance in closed-source scenarios, we construct a function-level evaluation benchmark through manual annotation. The annotation process involves three senior developers from our group, each with over five years of industrial experience, following four rules:
\begin{itemize}
    \item \textbf{Function Significance:} Selected functions must be integral to the daily development workflow within the codebase, representing real-world code completion challenges.
    \item \textbf{Context Selection:} Annotators manually identify relevant context and documentation (including line numbers and necessary explanations) based on their professional experience, simulating realistic code completion scenarios.
    \item \textbf{Difficulty Classification:} Each example is categorized as either `easy' or `hard' based on the complexity of the required completion, facilitating fine-grained analysis.
    \item \textbf{Quality Assurance:} All selected functions come from production systems that have undergone rigorous testing and are actively used in online environments. This ensures their correctness and practical value. After the initial annotation, all three developers perform cross-validation to ensure consistency and adherence to guidelines.
\end{itemize}

After three weeks of effort, we obtain a benchmark dataset containing 100 examples across seven domains. These domains cover essential enterprise development scenarios, ranging from remote procedure calls (client call), coroutine (connection, colib, and encoding), data storage operations (kv), message queue (mq), and utility functions (utils), representing common patterns in the software development of WeChat. Figure \ref{fig:benchmark} illustrates the distribution of examples across domains and difficulty levels.

\subsection{Retrieval Corpus Construction}
\label{sec:2.2}
We collect 1,669 internal projects as our retrieval database source. These projects span multiple business units and development cycles, providing a comprehensive codebase for retrieval. To ensure data quality, we filter out duplicate code snippets and standardize the code format. However, constructing a retrieval corpus from closed-source C++ projects presents several unique challenges:

\begin{itemize} 
    \item \textbf{File Segmentation:} C++ heavily relies on header files for dependency management, which often contain extensive object declarations and definitions \cite{mcintosh2016identifying}. Using entire files as retrieval units would result in excessively long retrieved segments. Meanwhile, applying sliding window approaches to split these files would likely fragment object contents and break their semantic integrity \cite{DBLP:conf/iclr/0008PWM0LSBSC24}.
    
    \item \textbf{Recursive Dependencies:} Header files frequently reference other header files, creating recursive dependency structures. This complexity prevents the direct application of dependency resolution approaches used in previous works \cite{24-Cheng,repofuse,liu2024graphcoder}.
    
    \item \textbf{Auto-generated Code:} Protocol Buffers (protobuf) is a language-agnostic data serialization format that allows developers to define structured data in proto files. While the protobuf compiler can automatically generate corresponding C++ files, these generated files contain numerous templates and predefined content that are irrelevant to the actual function logic \cite{hatledal2019language}.
    
    \item \textbf{Macro Specificity:} Unlike other object-oriented languages such as Python and Java, C++ extensively uses macro declarations, definitions, and implementations that play crucial roles in code functionality. These specific features require special handling during preprocessing.
\end{itemize}

To address these challenges, we develop a fine-grained preprocessing algorithm that extracts relevant objects from our internal codebase and uses these extracted objects as the basic units in our retrieval corpus. This granular organization ensures more precise and relevant retrievals while maintaining the semantic integrity of the code. Algorithm~\ref{alg:extract} presents our solution, which addresses each challenge as follows:

(1) To address the segmentation challenge, we extract class definitions and function definitions/declarations separately from cpp and header files, enabling fine-grained slicing and corpus construction. This is implemented as the main workflow in Algorithm~\ref{alg:extract}.

\begin{algorithm}[t]
\caption{Fine-grained Data Preprocessing}
\renewcommand{\algorithmicrequire}{\textbf{Input:}}
\renewcommand{\algorithmicensure}{\textbf{Output:}}
\begin{algorithmic}[1]
\label{alg:extract}
\REQUIRE Source file $f$, Processed headers set $H$
\ENSURE Function definitions $F_{def}$, declarations $F_{dec}$, class definitions $C$, or message definitions $M$
    \STATE Initialize $F_{def}, F_{dec}, C, M, F_{macro} \gets \emptyset$
    
    \IF{$f.type = \text{"proto"}$}
        \STATE $M \gets \text{ExtractProtoMessages}(f)$
        \RETURN $M$
    \ELSIF{$f.type = \text{"cpp"}$}
        \STATE // Extract elements from current file
        \STATE $C_f, F_{def}, F_{dec}, F_{macro} \gets \text{Extract}(f, \emptyset)$
        
        \STATE $headers \gets \text{GetRecursiveDependencies}(f)$
        \FOR{each $h \in headers$}
            \IF{$h \notin H$}
                \STATE // Recursively process dependent headers
                \STATE $C_h, F_{def\_h}, F_{dec\_h}, F_{macro\_h} \gets \text{Extract}(h, H)$
                \STATE $C \gets C \cup C_h$
                \STATE $F_{def} \gets F_{def} \cup F_{def\_h}$
                \STATE $F_{dec} \gets F_{dec} \cup F_{dec\_h}$
                \STATE $F_{macro} \gets F_{macro} \cup F_{macro\_h}$
                \STATE $H \gets H \cup \{h\}$
            \ENDIF
        \ENDFOR

        \STATE // Process macros
        \FOR{each $m \in F_{macro}$}
            \STATE $F_{def\_m}, F_{dec\_m} \gets \text{TransformMacro}(m)$
            \STATE $F_{def} \gets F_{def} \cup F_{def\_m}$
            \STATE $F_{dec} \gets F_{dec} \cup F_{dec\_m}$
        \ENDFOR
        
        \STATE // Remove redundant whitespace and comments
        \STATE $F_{def}, F_{dec}, C \gets \text{Format}(F_{def}, F_{dec}, C)$
        
        \RETURN $F_{def}, F_{dec}, C$
    \ELSE
        \STATE \textbf{Error: Unsupported file type!}
    \ENDIF
\end{algorithmic}
\end{algorithm}
    
(2) For recursive dependencies, we process all header files referenced in C++ source files recursively rather than only considering first-level dependencies. This process (lines 8-18 of Algorithm \ref{alg:extract}) ensures comprehensive dependency coverage while avoiding noise from unused dependencies.
    
(3) To handle auto-generated code issues, we first remove all the cpp and header files automatically generated from proto files. Considering that each protobuf message directly corresponds to a C++ class in the generated code, we design a specialized function \texttt{ExtractProtoMessages} (lines 2-4 of Algorithm \ref{alg:extract}) to extract message definitions directly from the more concise and structured protobuf files.
    
(4) For macro-specific features, we extract macro-related definitions and declarations from both cpp and header files, transforming them into function-like structures through pattern conversion. This transformation process is handled in lines 21-25 of Algorithm \ref{alg:extract}.

Specifically, a comprehensive retrieval corpus can be constructed from multiple C++ projects. 
The extraction process operates on both C++ source files (.cpp) and header files (.h), systematically identifying and extracting three types of background knowledge.
The set $F_{def}$ encompasses all function definitions, and the set $F_{dec}$ represents all function declarations, including both member functions and standalone functions. The set $C$ contains all class definitions existing in the codebase, representing the object-oriented structure of the system. The extraction process also operates on protobuf files (.proto). The set $M$ consists of protobuf message definitions, which are particularly important as they define the data structures used for communication and serialization.

\subsection{Identifier-based RAG}
Identifier-based retrieval-augmented generation aims to enhance code completion performance by incorporating the knowledge of relevant class, function, and protobuf message from the retrieval corpus. This section details the three main steps of this method: (1) index creation, (2) identifier extraction, and (3) prompt construction and code completion.

\subsubsection{Index Creation}
\label{sec:2.3.1}
To facilitate efficient retrieval, we construct an indexed codebase that enables quick lookup of background knowledge based on the identifier and search type:
\begin{equation}
    background\_knowledge = Lookup(identifier, type)
\end{equation}
This \textit{Lookup} function provides a service to access the background knowledge of specific objects, where $identifier$ refers to a string that uniquely identifies objects, such as the protobuf message name, class name, or function name. Type represents the search type, including protobuf message definition, function declaration, function definition, and class definition.

\subsubsection{Identifier Extraction}
\label{sec:2.3.2}
The second step leverages a powerful LLM to analyze the current code snippet and identify relevant and important references that require definition lookup. Through carefully designed prompts, the LLM understands the code context and extracts three sets of identifiers:
\begin{equation}
{M_{req}}, F_{req}, C_{req} = Need\_To\_Lookup(current\_code)
\end{equation}
where $M_{req}$ represents required protobuf message definitions, $F_{req}$ represents required function definitions and declarations, and $C_{req}$ represents required class definitions. Once the required references are identified, the corresponding background knowledge can be retrieved from the indexed codebases.

\subsubsection{Prompt Construction and Code Completion} 
The final step involves constructing specialized prompts for different types of background knowledge and finishing code completions. The process is formalized as:
\begin{equation}
\begin{aligned}
    generated\_code = LLM(&prompt\_template_{type}, \\ &knowledge_{type}, \\
    &current\_code)
\end{aligned}
\end{equation}
where $type$ has the same choices in Section \ref{sec:2.2}. In detail, we develop four distinct prompt templates to help LLMs understand different types of background knowledge.

\subsection{Similarity-based RAG}
Similarity-based retrieval augmented generation improves the performance of LLMs on code completion by providing code snippets similar to the current code. This process also includes three main steps: (1) index creation, (2) similar code retrieval, and (3) prompt construction and code completion.

\subsubsection{Index Creation}
Given that our primary focus is function completion, we only utilize function definitions as the retrieval source for the similarity-based RAG method. 
The construction of the retrieval corpus depends on the chosen similarity-based retrieval technique.
The indexing process of lexical retrieval is shown as follows:
\begin{equation}
     \{(term_i, tf_i, idf_i) | term_i \in F_{def}\}  \rightarrow lexical\_index
\end{equation}
where $tf_i$ represents the term frequency and $idf_i$ represents the inverse document frequency of each term in the corpus.
The indexing process of semantic retrieval is formalized as:
\begin{equation}
    \{encode(f) | f \in F_{def}\} \rightarrow semantic\_index
\end{equation}
where each function is encoded into a fixed-dimensional embedding space and stored in a vector database for efficient semantic search.

\subsubsection{Similar Code Retrieval}
In this step, the current code snippet serves as a query to retrieve similar code from the previously constructed retrieval corpus. The specific retrieval process depends on the chosen retrieval technique.

For lexical retrieval, the term frequencies of the current code are calculated, and the similarity is then computed based on the TF-IDF weights:

\begin{equation} 
similar\_code = argmax_{f \in F_{def}} \sum_{term \in query} TF\text{-}IDF(term, f) 
\end{equation}

For semantic retrieval, the current code is first encoded into an embedding with the same dimensionality as those in the retrieval corpus. The similarity is then measured using cosine similarity:
\begin{equation} 
\begin{aligned}
similar\_code = argmax_{f \in F_{def}} cos( &encode(query), \\
&encode(f))
\end{aligned}
\end{equation}

\subsubsection{Prompt Construction and Code Completion}
In this step, similar code snippets are integrated into the input of LLMs by prompt construction. Specifically, the retrieved similar code snippets are directly concatenated with the current code to construct the prompt:
\begin{equation}
\begin{aligned}
    generated\_code = LLM(&prompt\_template_{similar}, \\
    &similar\_code \oplus current\_code)
\end{aligned}
\end{equation}
where $\oplus$ represents the concatenation operation. 
In the prompt template, we encourage LLMs to complete the current code according to the similar code snippets.

\section{Experiment Study Setup}
\label{sec-3}
\subsection{Similarity-based Retrieval Techniques}
In our experimental evaluation, we employ five distinct similarity-based retrieval techniques, including one lexical technique (i.e., BM25\cite{BM25}) and four semantic techniques (i.e., CodeBERT \cite{codebert}, UniXcoder \cite{unixcoder}, CoCoSoDa \cite{cocosoda}, and GTE-Qwen \cite{gte-embedding}):

\textbf{BM25} \cite{BM25} extends the basic TF-IDF approach by incorporating document length normalization and non-linear term frequency scaling. In our implementation, we use BM25 to retrieve similar code snippets from a function corpus $F$. For a given code query, BM25 first tokenizes it into terms $\{t_1, t_2, ..., t_n\}$ and computes a relevance score for each function $f \in F$ as:
\begin{equation}
    Score(query, f) = \sum^{n}_{i=1}(IDF_i \cdot TF_{mod}(t_i, f))
\end{equation}
The IDF component is calculated similarly to the standard TF-IDF approach, but with smoothing factors:
\begin{equation}
    IDF_i = log\frac{N-df_i+0.5}{df_i+0.5}
\end{equation}
where $N$ is the total number of functions in the corpus and $df_i$ is the number of functions containing term $t_i$.
The modified term frequency component $TF_{mod}$ introduces saturation and length normalization:
\begin{equation}
    TF_{mod}(t_i, f) = \frac{tf_i \cdot (k+1)}{tf_i + k \cdot (1-b+b \cdot \frac{len_f}{len_{avg}})}
\end{equation}
where $tf_i$ is the frequency of term $t_i$ in function $f$, $len_f$ is the length of function $f$, and $len_{avg}$ is the average function length in the corpus. Parameters $k$ and $b$ control term frequency scaling and length normalization, respectively.

\textbf{CodeBERT} \cite{codebert} 
is built upon the RoBERTa-base and pre-trained on both natural language and programming language corpus. It employs two types of objective modeling tasks, including Masked Language Modeling (MLM) and Replaced Token Detection (RTD), which enables the model to learn general representations.

\textbf{UniXcoder} \cite{unixcoder} 
is pre-trained using three objectives: masked language modeling, unidirectional language modeling, and denoising tasks. Additionally, UniXcoder incorporates two pre-training strategies: (1) a multi-modal contrastive learning approach that leverages Abstract Syntax Tree (AST) to enhance code representations, and (2) a cross-modal generation task utilizing code comments to align embeddings across different programming languages.

\textbf{CoCoSoDa} \cite{cocosoda}, which has the same architecture as UniXcoder, enhances code representation through momentum contrastive learning. Specifically, the model employs dynamic data augmentation and negative sampling strategies
to learn more robust and discriminative code embeddings.

\textbf{GTE-Qwen} \cite{gte-embedding}, a member of the GTE (General Text Embedding) family, is a decode-only retrieval model based on Qwen2 \cite{qwen2}. 
Different from the general Qwen model, GTE-Qwen incorporates bidirectional attention mechanisms to enrich contextual representations, making it particularly effective for multilingual understanding and retrieval tasks. We utilize GTE-Qwen2-1.5B-instruct in our experiment.

\subsection{Large Language Models}

To comprehensively evaluate code generation capabilities across different model scales, we conduct experiments with 26 open-source LLMs, ranging from 0.5B to 671B parameters. 

Our evaluation encompasses both code-specialized LLMs and general-purpose LLMs. For code-specialized models, we select several prominent and emerging model series, including Qwen-Coder \cite{qwen2.5coder}, Deepseek-Coder \cite{deepseekllm, deepseekcoderv2}, CodeLlama \cite{codellama}, Yi-Coder \cite{yi-coder}, OpenCoder \cite{opencoder}, and Codestral \cite{Codestral}. For each series, we incorporate their latest versions across different parameter scales to ensure comprehensive coverage. 
To complement our evaluation and fill the gaps in model scale coverage, we also include several general-purpose LLMs, namely Llama-3.2-1B-Instruct, Llama-3.2-3B-Instruct \cite{Llama-3.2}, Llama-3.1-8B-Instruct \cite{Llama-3.1}, Llama-3.3-70B-Instruct \cite{Llama-3.3}, Qwen2.5-72B-Instruct \cite{qwen2.5}, as well as DeepSeek-V2.5 \cite{deepseekcoderv2} and DeepSeek-V3 \cite{deepseekv3}. 

\subsection{Metrics}

\textbf{CodeBLEU (CB)} \cite{codebleu} 
extends the BLEU metric \cite{bleu} by incorporating additional factors to capture the structural and semantic aspects of code.
Specifically, CodeBLEU is formally defined as a weighted combination of four factors:
\begin{equation}
\begin{aligned}
CodeBLEU = \alpha \cdot N\text{-}gram_{match} &+ \beta \cdot N\text{-}gram_{weighted}\;+\\
\gamma \cdot Similarity_{AST} &+ \delta \cdot Similarity_{DF}
\end{aligned}
\end{equation}
where $N\text{-}gram_{match}$ measures the overlap of code tokens between the generated and reference code. $N\text{-}gram_{weighted}$ applies different weights to tokens based on their importance in context. $Similarity_{AST}$ evaluates the structural similarity by calculating the syntactic AST matching score, and $Similarity_{DF}$ assesses the semantic equivalence through data flow analysis. The coefficients $\alpha$, $\beta$, $\gamma$, and $\delta$ control the relative contribution of each factor, allowing for flexible adaptation to different evaluation scenarios. In our experiments, all four coefficients are set to 0.25.

\textbf{Edit Similarity (ES)} measures the minimal number of token-level operations required to transform one string into another, normalized by the length of the longer string. Formally, given a generated code snippet $c_g$ and its reference $c_r$, the edit similarity is calculated as:
\begin{equation}
    ES(c_g, c_r) = 1 - \frac{EditDistance(c_g, c_r)}{\max(|c_g|, |c_r|)}
\end{equation}
where $EditDistance(c_g, c_r)$ refers to the Levenshtein distance between two strings, counting the minimum number of single-token insertions, deletions, or substitutions needed to transform $c_g$ into $c_r$. The resulting score ranges from 0 to 1, where 1 indicates identical sequences and 0 represents completely different strings. 

For better readability, we present the two metrics as percentage scores (multiplied by 100) in our experimental results.

\begin{table*}[htbp]
\centering
\caption{Performance comparison of LLMs with different RAG methods. The metrics showing improvements over the base model are highlighted by \graycell{gray}. The best performance metrics within each RAG category are marked in \textbf{bold}.}
\label{tab:tab1}
\resizebox{\textwidth}{!}{
\begin{tabular}{cc|cccc|ccccc}
\Xhline{1.2pt}
\multirow{3}{*}{\textbf{Model}} &  & \multicolumn{4}{c|}{\textbf{Identifier-Based RAG}} & \multicolumn{5}{c}{\textbf{Similarity-Based RAG}}\\
& \textbf{base} & \textbf{msg-def.} & \textbf{class-def.} & \textbf{func-dec.} & \textbf{func-def.} & \textbf{BM25} & \textbf{CodeBERT} & \textbf{UniXcoder} & \textbf{CoCoSoDa} & \textbf{GTE-Qwen} \\
 & CB/ES & CB/ES & CB/ES & CB/ES & CB/ES & CB/ES & CB/ES & CB/ES & CB/ES & CB/ES \\
\hline 
\hline 
\multicolumn{11}{c}{0.5B+ LLM} \\
\hline 
Qwen2.5-Coder-0.5B-Instruct & 27.57/41.54 & 24.39/\textbf{38.31} & 24.25/37.36  & 21.38/36.09 & \textbf{26.72}/37.71 & \graycell{31.43}/41.25 & 23.96/36.92 & \graycell{29.98}/39.43 & \graycell{30.39}/38.68 & \graycell{\textbf{34.46}}/\graycell{\textbf{41.67}} \\
\hline
\multicolumn{11}{c}{1B+ LLMs} \\
\hline
OpenCoder-1B-Instruct & 23.28/29.83 & \textbf{23.27}/\graycell{\textbf{31.06}} & 16.39/23.49  & 20.52/\graycell{30.28} & 23.10/\graycell{30.40} & \graycell{\textbf{27.63}}/\graycell{\textbf{32.45}} & \graycell{23.60}/\graycell{30.07} & 21.22/\graycell{30.79} & 22.67/29.47 & 22.12/28.13 \\
Llama-3.2-1B-Instruct & 25.78/32.40 & 24.50/\textbf{31.63} & 20.33/29.50  & 23.64/31.56 & \textbf{25.75}/29.40 & \graycell{30.18}/\graycell{32.59} & 24.30/29.86 & 25.39/30.02 & 25.12/30.06 & \graycell{\textbf{30.93}}/\graycell{\textbf{32.64}} \\
DS-Coder-1.3B-Instruct & 21.85/35.22 & \graycell{22.08}/\graycell{\textbf{36.08}} & \multicolumn{1}{p{4em}}{\graycell{\textbf{24.85}}/33.17} &  \graycell{22.20}/\graycell{35.52} & \multicolumn{1}{p{4em}|}{\graycell{23.62}/34.48} & \graycell{31.09}/\graycell{\textbf{39.25}} & 20.53/33.83 & \graycell{25.01}/\graycell{35.54} & \graycell{27.76}/\graycell{35.49} & \graycell{\textbf{34.45}}/\graycell{37.06} \\
Qwen2.5-Coder-1.5B-Instruct & 24.42/42.83 & \graycell{35.05}/\graycell{\textbf{51.13}} & \graycell{31.82}/\graycell{46.10} & \graycell{32.00}/\graycell{46.99} & \graycell{\textbf{37.28}}/\graycell{50.77} & \graycell{\textbf{47.78}}/\graycell{\textbf{57.62}} & 17.33/34.26 & \graycell{35.82}/\graycell{50.22} & \graycell{41.42}/\graycell{50.85} & \graycell{46.69}/\graycell{56.04} \\
Yi-Coder-1.5B-Chat & 19.56/31.69  & \graycell{\textbf{21.66}}/\graycell{34.41} & 18.67/31.52 & \graycell{21.06}/\graycell{34.47} & \graycell{20.97}/\graycell{\textbf{34.37}} & \graycell{\textbf{30.17}}/\graycell{\textbf{38.38}} & 18.45/\graycell{34.02} & \graycell{23.86}/\graycell{34.07} & \graycell{24.23}/\graycell{34.98} & \graycell{29.06}/\graycell{36.40} \\
\hline
\multicolumn{11}{c}{3B+ LLMs} \\
\hline
Llama-3.2-3B-Instruct & 32.79/45.74 & 32.23/44.82 & \multicolumn{1}{p{4em}}{\graycell{32.97}/39.71} &  \graycell{33.64}/45.13 & \graycell{\textbf{36.02}}/\graycell{\textbf{45.83}} & \multicolumn{1}{p{4em}}{\graycell{45.42}/\graycell{47.37}} & \graycell{34.15}/44.66 & \multicolumn{1}{p{4em}}{\graycell{38.25}/44.80} &\multicolumn{1}{p{4em}}{\graycell{41.19}/45.62} & \graycell{\textbf{49.04}}/\graycell{\textbf{51.07}} \\
Qwen2.5-Coder-3B-Instruct & 15.41/38.15 & \graycell{17.07}/\graycell{39.08} & \graycell{16.99}/\graycell{38.72}  & 14.45/34.97 & \graycell{\textbf{19.86}}/\graycell{\textbf{40.62}} & \graycell{27.12}/\graycell{44.48} & \graycell{18.25}/\graycell{40.82} & \graycell{17.57}/\graycell{38.52} & \graycell{24.89}/\graycell{42.88} & \graycell{\textbf{30.63}}/\graycell{\textbf{48.48}} \\
\hline
\multicolumn{11}{c}{7B+ LLMs} \\
\hline
DS-Coder-7B-Instruct-v1.5 & 33.24/49.54 & 32.59/48.35 & 27.75/40.52  & \graycell{34.19}/49.49 & \graycell{\textbf{37.62}}/\graycell{\textbf{50.11}} & \graycell{43.21}/\graycell{\textbf{55.20}} & \multicolumn{1}{p{4em}}{\graycell{34.05}/46.67} & \graycell{37.75}/\graycell{50.37} & \multicolumn{1}{p{4em}}{\graycell{35.12}/47.13} & \graycell{\textbf{44.44}}/\graycell{52.36} \\
Qwen2.5-Coder-7B-Instruct & 33.00/50.27 & \graycell{\textbf{34.60}}/\graycell{51.12} &\graycell{33.60}/\graycell{\textbf{51.73}} & 30.14/48.73 & \graycell{34.01}/\graycell{51.10} & \graycell{45.16}/\graycell{59.36} & 32.44/\graycell{52.13} & \graycell{38.36}/\graycell{54.75} & \graycell{44.23}/\graycell{60.96} & \graycell{\textbf{49.03}}/\graycell{\textbf{62.66}} \\
Llama-3.1-8B-Instruct & 34.02/46.07 & \multicolumn{1}{p{4em}}{\graycell{36.42}/45.33} & \graycell{36.86}/\graycell{46.34}  & \graycell{35.55}/45.86 & \graycell{\textbf{39.64}}/\graycell{\textbf{49.35}} & \graycell{49.80}/\graycell{54.39} & \multicolumn{1}{p{4em}}{\graycell{35.07}/45.55} & \graycell{40.01}/\graycell{49.00} & \graycell{46.22}/\graycell{51.65} & \graycell{\textbf{53.47}}/\graycell{\textbf{55.40}} \\
OpenCoder-8B-Instruct & 29.69/31.42 & \multicolumn{1}{p{4em}}{\graycell{30.28}/29.96} & 28.65/29.30 & \graycell{32.89}/29.42 & \multicolumn{1}{p{4em}|}{\graycell{\textbf{37.35}}/\textbf{30.16}} & \graycell{\textbf{42.45}}/\graycell{32.17} & \graycell{29.70}/27.71 & \multicolumn{1}{p{4em}}{\graycell{34.37}/30.08} & \multicolumn{1}{p{4em}}{\graycell{37.15}/30.04} & \graycell{41.38}/\graycell{\textbf{32.28}} \\
Yi-Coder-9B-Chat & 33.78/47.11 & 33.69/45.25 & 32.00/43.63  & \graycell{34.03}/46.07 & \multicolumn{1}{p{4em}}{\graycell{\textbf{35.49}}/\textbf{46.98}} & \graycell{\textbf{51.66}}/\graycell{\textbf{56.14}} & \multicolumn{1}{p{4em}}{\graycell{34.49}/46.57} & \graycell{40.28}/\graycell{49.99} & \graycell{42.87}/\graycell{50.49} & \graycell{49.59}/\graycell{55.73} \\
\hline
\multicolumn{11}{c}{13B+ LLMs} \\
\hline
CodeLlama-13B-Instruct & 26.81/35.52 & \textbf{24.09}/30.78 & 21.05/28.35 &  24.01/\textbf{31.32} & 21.82/28.75 & \multicolumn{1}{p{4em}}{\graycell{\textbf{28.00}}/30.51} & 23.23/29.54 & 23.84/28.66 & 24.63/29.24 & 26.58/\textbf{30.70} \\
Qwen2.5-Coder-14B-Instruct & 29.79/48.56 & \graycell{33.43}/\graycell{52.33} & 28.14/47.54  & \graycell{33.86}/\graycell{\textbf{53.56}} & \graycell{\textbf{35.02}}/\graycell{53.23} & \graycell{46.07}/\graycell{59.97} & \graycell{35.06}/\graycell{53.82} & \graycell{35.75}/\graycell{52.25} & \graycell{43.01}/\graycell{57.50} & \graycell{5\textbf{1.12}}/\graycell{\textbf{61.96}} \\
DS-Coder-V2-Lite-Instruct-16B/2.4B & 34.72/51.01 & \graycell{35.31}/\graycell{\textbf{51.33}} & 33.93/46.48 &  33.49/49.68 & \multicolumn{1}{p{4em}|}{\graycell{\textbf{39.34}}/48.15} & \graycell{51.45}/\graycell{57.87} & 34.23/49.10 & \multicolumn{1}{p{4em}}{\graycell{40.83}/50.75} & \graycell{45.60}/\graycell{55.25} & \graycell{\textbf{54.91}}/\graycell{\textbf{57.98}} \\
\hline
\multicolumn{11}{c}{20B+ LLMs} \\
\hline
CodeStral-22B-v0.1 & 34.12/55.25 & \multicolumn{1}{p{4em}}{\graycell{36.13}/54.52} & \multicolumn{1}{p{4em}}{\graycell{\textbf{36.86}}/54.64}  & \multicolumn{1}{p{4em}}{\graycell{34.51}/\textbf{55.03}} & \multicolumn{1}{p{4em}|}{\graycell{36.11}/54.99} & \graycell{47.28}/\graycell{\textbf{60.93}} & \multicolumn{1}{p{4em}}{\graycell{36.17}/54.65} & \multicolumn{1}{p{4em}}{\graycell{36.16}/52.66} & \graycell{43.94}/\graycell{57.05} & \graycell{\textbf{49.37}}/\graycell{60.80} \\
Qwen2.5-Coder-32B-Instruct & 38.05/57.89 & \graycell{38.91}/\graycell{58.84} & \graycell{40.54}/\graycell{59.52} & 37.24/57.46 & \graycell{\textbf{42.23}}/\graycell{\textbf{60.44}} & \graycell{55.76}/\graycell{68.89} & \graycell{38.67}/\graycell{59.28} & \graycell{44.16}/\graycell{61.13} & \graycell{49.50}/\graycell{65.54} & \graycell{\textbf{60.79}}/\graycell{\textbf{71.34}} \\
DS-Coder-33B-Instruct & 28.48/45.20 & \graycell{31.45}/\graycell{\textbf{48.34}} & 24.47/39.66 & \graycell{30.36}/\graycell{47.99} & \graycell{\textbf{32.25}}/\graycell{46.74} & \graycell{38.91}/\graycell{\textbf{50.19}} & \multicolumn{1}{p{4em}}{\graycell{29.37}/44.48} & \graycell{34.02}/\graycell{47.96} & \graycell{34.50}/\graycell{46.23} & \graycell{\textbf{39.40}}/\graycell{47.43} \\
CodeLlama-34B-Instruct & 26.51/37.93 & 23.60/34.58 & 19.15/32.23 & \textbf{23.88}/\textbf{35.33} & 22.37/33.45 & \multicolumn{1}{p{4em}}{\graycell{27.53}/36.48} & 21.36/32.30 & 18.01/31.58 & 22.12/33.93 & \multicolumn{1}{p{4em}}{\graycell{\textbf{28.35}}/\textbf{37.42}} \\
\hline
\multicolumn{11}{c}{70B+ LLMs} \\
\hline
CodeLlama-70B-Instruct & 22.50/33.10 & \textbf{17.88}/\textbf{28.95} & 12.06/17.85 &  12.59/21.99 & 11.60/17.39 & \textbf{19.17}/\textbf{26.95} & 13.65/20.57 & 15.23/21.34 & 16.05/20.30 & 16.26/20.70 \\
Llama-3.3-70B-Instruct & 34.14/53.04 & \multicolumn{1}{p{4em}}{\graycell{35.15}/52.53} & \graycell{36.21}/\graycell{53.30} & \graycell{36.92}/\graycell{55.30}   & \graycell{\textbf{40.78}}/\graycell{\textbf{58.39}} & \graycell{50.21}/\graycell{62.37} & \graycell{36.93}/\graycell{54.84} & \graycell{39.33}/\graycell{55.02} & \graycell{45.53}/\graycell{58.86} & \graycell{\textbf{52.64}}/\graycell{\textbf{64.60}} \\
Qwen2.5-72B-Instruct & 37.03/54.71 & \graycell{38.66}/\graycell{55.97} & \graycell{38.07}/\graycell{55.41} & \graycell{38.72}/\graycell{56.86}  & \graycell{\textbf{41.90}}/\graycell{\textbf{59.02}} & \graycell{50.21}/\graycell{62.45} & \graycell{37.89}/\graycell{56.74} & \graycell{45.57}/\graycell{60.38} & \graycell{48.16}/\graycell{61.96} & \graycell{\textbf{56.05}}/\graycell{\textbf{65.35}} \\
\hline
\multicolumn{11}{c}{200B+ LLMs} \\
\hline
DS-Coder-V2-Instruct-236B/21B & 33.26/54.92 & \graycell{38.35}/\graycell{57.95} & \graycell{37.63}/\graycell{55.54} &  \graycell{38.40}/\graycell{58.50} & \graycell{\textbf{43.52}}/\graycell{\textbf{62.29}} & \graycell{53.72}/\graycell{\textbf{69.27}} & \graycell{34.27}/\graycell{55.73} & \graycell{44.56}/\graycell{61.60} & \graycell{48.58}/\graycell{63.33} & \graycell{\textbf{55.92}}/\graycell{69.06} \\
DeepSeek-V2.5-236B/21B & 33.50/54.32 & \graycell{38.26}/\graycell{56.55} &  \graycell{36.43}/\graycell{55.16}  & \graycell{38.25}/\graycell{58.74} & \graycell{\textbf{41.81}}/\graycell{\textbf{61.82}} & \graycell{\textbf{55.67}}/\graycell{\textbf{69.18}} & 32.67/53.75 &   \graycell{44.70}/\graycell{61.69} &  \graycell{48.16}/\graycell{63.85} &  \graycell{55.29}/\graycell{68.21}  \\
DeepSeek-V3-671B/37B & 35.23/54.85  & \graycell{39.04}/\graycell{58.54} & \graycell{37.74}/\graycell{57.53} & \graycell{37.51}/\graycell{58.63} & \graycell{\textbf{42.24}}/\graycell{\textbf{61.75}} &  \graycell{55.14}/\graycell{68.55} & \graycell{38.13}/\graycell{58.64} & \graycell{44.75}/\graycell{62.27} & \graycell{50.43}/\graycell{65.40} & \graycell{\textbf{60.28}}/\graycell{\textbf{73.11}} \\
\Xhline{1.2pt}
\end{tabular}
}
\end{table*}

\subsection{Implementation Details}
\subsubsection{Data Preprocessing}
We implement the data preprocessing process following the algorithm outlined in Section \ref{sec:2.2}. For the \texttt{Extract} function,
we employ the tree-sitter and s-expression pattern matching to efficiently extract the required patterns in constant time. 
Different from processing C++ source files and header files with tree-sitter, the protobuf files cannot be parsed into ASTs directly. To overcome this limitation, we carefully study the official protobuf documentation and design a regular expression-based approach to extract Message definitions. 

For macro transformations, we develop a systematic approach based on three key components: macro parameters, internal logic, and return types. The macros can be converted into function-like structures by using macro names as function names and macro parameters as function parameters. Macros without internal implementation logic are transformed into function declarations, while those with both internal logic and return types are converted into function definitions.

\begin{table*}[htbp]
\centering
\caption{Comparative analysis of retrieval techniques in similarity-based RAG. The best performance metrics within each retrieval technique are highlighted by \graycell{gray}. ``Incomplete'' denotes using partial code context as retrieval query, while ``Complete'' represents using the entire code snippet for retrieval. The best performance metrics with incomplete code snippets as queries for each LLM are marked in \textbf{bold}.}
\label{tab:tab2}
\resizebox{\textwidth}{!}{
\begin{tabular}{c|cc|cc|cc|cc|cc}
\Xhline{1.2pt}
\multirow{3}{*}{\textbf{Model}}& \multicolumn{2}{c|}{\textbf{BM25}} & \multicolumn{2}{c|}{\textbf{CodeBERT}} & \multicolumn{2}{c|}{\textbf{UniXcoder}} & \multicolumn{2}{c|}{\textbf{CoCoSoDa}} & \multicolumn{2}{c}{\textbf{GTE-Qwen}} \\
 & \textbf{Incomplete} & \textbf{Complete} & \textbf{Incomplete} & \textbf{Complete} & \textbf{Incomplete} & \textbf{Complete} & \textbf{Incomplete} & \textbf{Complete} & \textbf{Incomplete} & \textbf{Complete} \\
 & CB/ES & CB/ES & CB/ES & CB/ES & CB/ES & CB/ES & CB/ES & CB/ES & CB/ES & CB/ES \\
\hline
\hline
\multicolumn{11}{c}{0.5B+ LLM} \\
\hline
Qwen2.5-Coder-0.5B-Instruct & 31.43/41.25 & \graycell{36.40}/\graycell{43.48} & 23.96/\graycell{36.92} & \graycell{24.00}/36.70 & 29.98/39.43 & \graycell{34.16}/\graycell{41.65} & 30.39/38.68 & \graycell{34.27}/\graycell{41.12} & \graycell{\textbf{34.46}}/\graycell{\textbf{41.67}} & 33.77/39.62 \\
\hline
\multicolumn{11}{c}{1B+ LLMs} \\
\hline
OpenCoder-1B-Instruct & \textbf{27.63}/\graycell{\textbf{32.45}} & \multicolumn{1}{p{4em}|}{\graycell{31.81}/32.31} & \multicolumn{1}{p{4em}}{\whitecell{23.60}/\graycell{30.07}} & \graycell{23.68}/29.82 & 21.22/30.79 & \graycell{26.89}/\graycell{31.65} & 22.67/\graycell{29.47} & \multicolumn{1}{p{4em}|}{\graycell{27.97}/28.32} & 22.12/28.13 & \multicolumn{1}{p{4em}}{\graycell{23.87}/\graycell{28.18}} \\
Llama-3.2-1B-Instruct & 30.18/32.59 & \graycell{36.75}/\graycell{33.68} & \multicolumn{1}{p{4em}}{\graycell{24.30}/\graycell{29.86}} & 22.92/28.84 & 25.39/30.02 & \graycell{30.02}/\graycell{31.76} & 25.12/30.06 & \graycell{29.66}/\graycell{31.82} & \multicolumn{1}{p{4em}}{\graycell{\textbf{30.93}}/\textbf{32.64}} & 30.58/\graycell{32.71} \\
DS-Coder-1.3B-Instruct & 31.09/\textbf{39.25} & \graycell{37.30}/\graycell{40.53} & \multicolumn{1}{p{4em}}{\graycell{20.53}/33.83} & \whitecell{20.19}/\graycell{34.15} & 25.01/35.54 & \graycell{31.54}/\graycell{39.92} & 27.76/35.49 & \graycell{33.00}/\graycell{37.49} & \multicolumn{1}{p{4em}}{\graycell{\textbf{34.45}}/\graycell{37.06}} & 33.64/36.04 \\
Qwen2.5-Coder-1.5B-Instruct & \textbf{47.78}/\textbf{57.62} & \graycell{51.84}/\graycell{59.39} & \multicolumn{1}{p{4em}}{\whitecell{17.33}/\graycell{34.26}} & \graycell{17.51}/33.72 & 35.82/50.22 & \graycell{42.89}/\graycell{53.25} & 41.42/50.85 & \graycell{44.60}/\graycell{51.68} & \graycell{46.69}/\graycell{56.04} & 44.52/54.12 \\
Yi-Coder-1.5B-Chat & \textbf{30.17}/\graycell{\textbf{38.38}} & \multicolumn{1}{p{4em}|}{\graycell{34.57}/38.31} & \graycell{18.45}/\graycell{34.02} & 17.78/32.87 & 23.86/34.07 & \graycell{32.23}/\graycell{38.37} & 24.23/34.98 & \graycell{28.11}/\graycell{37.00} & \graycell{29.06}/\graycell{36.40} & 28.57/36.28 \\
\hline
\multicolumn{11}{c}{3B+ LLMs} \\
\hline
Llama-3.2-3B-Instruct & 45.42/47.37 & \graycell{50.74}/\graycell{49.82} & \graycell{34.15}/\graycell{44.66} & 33.72/43.32 & 38.25/44.80 & \graycell{46.56}/\graycell{50.82} & 41.19/45.62 & \graycell{44.50}/\graycell{47.80} & \graycell{\textbf{49.04}}/\graycell{\textbf{51.07}} & 47.20/49.27 \\
Qwen2.5-Coder-3B-Instruct & 27.12/44.48 & \graycell{31.59}/\graycell{47.73} & \graycell{18.25}/\graycell{40.82} & 17.98/40.27 & 17.57/38.52 & \graycell{27.05}/\graycell{45.56} & 24.89/42.88 & \graycell{27.94}/\graycell{45.05} & \textbf{30.63}/\graycell{\textbf{48.48}} & \graycell{30.90}/47.80 \\
\hline
\multicolumn{11}{c}{7B+ LLMs} \\
\hline
DS-Coder-7B-Instruct-v1.5 & 43.21/\textbf{55.20} & \graycell{48.65}/\graycell{58.02} & 34.05/\graycell{46.67} & \multicolumn{1}{p{4em}}{\graycell{34.12}/46.48} & 37.75/50.37 & \graycell{45.00}/\graycell{55.39} & 35.12/47.13 & \graycell{37.86}/\graycell{48.09} & \textbf{44.44}/52.36 & \graycell{45.63}/\graycell{53.78} \\
Qwen2.5-Coder-7B-Instruct & 45.16/59.36 & \graycell{52.92}/\graycell{64.13} & 32.44/52.13 & \graycell{32.83}/\graycell{52.68} & 38.36/54.75 & \graycell{46.75}/\graycell{60.22} & 44.23/\graycell{60.96} & \multicolumn{1}{p{4em}|}{\graycell{46.60}/60.95} & \graycell{\textbf{49.03}}/\graycell{\textbf{62.66}} & 47.73/61.05 \\
Llama-3.1-8B-Instruct & 49.80/54.39 & \graycell{57.00}/\graycell{60.19} & 35.07/45.55 & \graycell{35.53}/\graycell{46.13} & 40.01/49.00 & \graycell{50.34}/\graycell{54.69} & 46.22/\graycell{51.65} & \multicolumn{1}{p{4em}|}{\graycell{48.96}/51.62} & \graycell{\textbf{53.47}}/\graycell{\textbf{55.40}} & 52.89/54.27 \\
OpenCoder-8B-Instruct & \textbf{42.45}/32.17 & \graycell{47.21}/\graycell{32.98} & \graycell{29.70}/\graycell{27.71} & 29.28/27.15 & 34.37/30.08 & \graycell{41.11}/\graycell{31.36} & 37.15/30.04 & \graycell{40.96}/\graycell{31.52} & \multicolumn{1}{p{4em}}{\whitecell{41.38}/\graycell{\textbf{32.28}}} & \graycell{41.97}/31.84 \\
Yi-Coder-9B-Chat & \textbf{51.66}/\textbf{56.14} & \graycell{56.53}/\graycell{58.91} & \whitecell{34.49}/\graycell{46.57} & \graycell{34.50}/45.80 & 40.28/49.99 & \graycell{48.88}/\graycell{54.10} & 42.87/50.49 & \graycell{48.23}/\graycell{54.00} & \graycell{49.59}/\graycell{55.73} & 47.92/52.65 \\
\hline
\multicolumn{11}{c}{13B+ LLMs} \\
\hline
CodeLlama-13B-Instruct & \textbf{28.00}/30.51 & \graycell{34.15}/\graycell{31.72} & \graycell{23.23}/\whitecell{29.54} & 22.88/\graycell{29.69} & 23.84/28.66 & \graycell{29.51}/\graycell{30.45} & 24.63/29.24 & \graycell{27.89}/\graycell{29.69} & 26.58/\graycell{\textbf{30.70}} & \graycell{28.10}/30.02 \\
Qwen2.5-Coder-14B-Instruct & 46.07/59.97 & \graycell{52.16}/\graycell{63.72} & \graycell{35.06}/\graycell{53.82} & 33.87/53.19 & 35.75/52.25 & \graycell{46.72}/\graycell{60.36} & 43.01/57.50 & \graycell{46.39}/\graycell{59.66} & \graycell{\textbf{51.12}}/\graycell{\textbf{61.96}} & 49.28/60.16 \\
DS-Coder-V2-Lite-Instruct-16B/2.4B & 51.45/57.87 & \graycell{57.67}/\graycell{62.19} & 34.23/49.10 & \graycell{34.65}/\graycell{50.20} & 40.83/50.75 & \graycell{49.35}/\graycell{55.13} & 45.60/55.25 & \graycell{49.64}/\graycell{56.08} & \graycell{\textbf{54.91}}/\graycell{\textbf{57.98}} & 53.50/56.00 \\
\hline
\multicolumn{11}{c}{20B+ LLMs} \\
\hline
CodeStral-22B-v0.1 & 47.28/\textbf{60.93} & \graycell{52.46}/\graycell{65.11} & \graycell{36.17}/\graycell{54.65} & 36.02/52.65 & 36.16/52.66 & \graycell{48.87}/\graycell{60.91} & 43.94/57.05 & \graycell{47.87}/\graycell{58.87} & \multicolumn{1}{p{4em}}{\graycell{\textbf{49.37}}/60.80} & 49.07/\graycell{61.02} \\
Qwen2.5-Coder-32B-Instruct & 55.76/68.89 & \graycell{61.98}/\graycell{72.38} & \graycell{38.67}/\graycell{59.28} & 38.42/58.77 & 44.16/61.13 & \graycell{54.97}/\graycell{69.16} & 49.50/65.54 & \graycell{52.63}/\graycell{66.52} & \graycell{\textbf{60.79}}/\graycell{\textbf{71.34}} & 59.75/70.25 \\
DS-Coder-33B-Instruct & 38.91/\textbf{50.19} & \graycell{42.32}/\graycell{53.85} & \graycell{29.37}/\graycell{44.48} & 28.89/43.69 & 34.02/47.96 & \graycell{42.06}/\graycell{52.79} & 34.50/46.23 & \graycell{37.40}/\graycell{48.39} & \textbf{39.40}/47.43 & \graycell{40.84}/\graycell{48.89} \\
CodeLlama-34B-Instruct & 27.53/36.48 & \graycell{32.47}/\graycell{37.83 }& 21.36/32.30 & \graycell{22.24}/\graycell{33.61} & 18.01/31.58 & \graycell{23.71}/\graycell{34.64} & 22.12/33.93 & \graycell{24.00}/\graycell{35.15} & \textbf{28.35}/\textbf{37.42} & \graycell{28.44}/\graycell{37.55} \\
\hline
\multicolumn{11}{c}{70B+ LLMs} \\
\hline
CodeLlama-70B-Instruct & \textbf{19.17}/\textbf{26.95} & \graycell{25.13}/\graycell{28.45} & \whitecell{13.65}/\graycell{20.57} & \graycell{13.97}/20.33 & 15.23/21.34 & \graycell{23.49}/\graycell{26.13} & 16.05/20.30 & \graycell{19.53}/\graycell{21.15} & 16.26/20.70 & \graycell{16.88}/\graycell{20.88} \\
Llama-3.3-70B & 50.21/62.37 & \graycell{58.39}/\graycell{66.70} & \graycell{36.93}/\graycell{54.84} & 36.24/54.61 & 39.33/55.02 & \graycell{49.15}/\graycell{62.24} & 45.53/58.86 & \graycell{50.09}/\graycell{61.58} & \graycell{\textbf{52.64}}/\graycell{\textbf{64.60}} & 52.03/62.48 \\
Qwen2.5-72B & 50.21/62.45 & \graycell{56.66}/\graycell{66.74} & \graycell{37.89}/\graycell{56.74} & 37.29/55.65 & 45.57/60.38 & \graycell{55.84}/\graycell{67.22} & 48.16/61.96 & \graycell{52.24}/\graycell{63.94} & \textbf{56.05}/\textbf{65.35} & \graycell{56.81}/\graycell{65.75} \\
\hline
\multicolumn{11}{c}{200B+ LLMs} \\
\hline
DS-Coder-V2-Instruct-236B/21B & 53.72/\textbf{69.27} & \graycell{61.65}/\graycell{72.39} & 34.27/\graycell{55.73} & \multicolumn{1}{p{4em}}{\graycell{34.54}/54.55} & 44.56/61.60 & \graycell{55.81}/\graycell{70.07} & 48.58/63.33 & \graycell{51.99}/\graycell{65.95} & \textbf{55.92}/\graycell{69.06} & \multicolumn{1}{p{4em}}{\graycell{56.63}/68.51} \\
DeepSeek-V2.5-236B/21B & \textbf{55.67}/\textbf{69.18} & \graycell{60.92}/\graycell{72.47} & 32.67/53.75 & \graycell{35.38}/\graycell{54.09}  & 44.70/61.69  & \graycell{55.00}/\graycell{70.14}  & 48.16/63.85  & \graycell{51.61}/\graycell{66.04} & 55.29/68.21 & \graycell{57.80}/\graycell{69.15} \\
DeepSeek-V3-671B/37B & 55.14/68.55 & \graycell{63.63}/\graycell{73.70} & \graycell{38.13}/58.64 & \whitecell{38.00}/\graycell{58.91} & 44.75/62.27 & \graycell{57.81}/\graycell{72.40} & 50.43/65.40 & \graycell{55.93}/\graycell{69.14} & \graycell{\textbf{60.28}}/\graycell{\textbf{73.11}} & 58.85/71.02  	 \\
\Xhline{1.2pt}
\end{tabular}
}
\end{table*}

\subsubsection{Retrieval Service}

For identifier-based RAG, we implement a retrieval service that provides specific background knowledge based on identifiers and retrieval types.
Specifically, we adopt Tantivy\footnote{\url{https://github.com/quickwit-oss/tantivy}}, an efficient full-text search engine library, to construct separate retrieval databases for different retrieval types (e.g., protobuf message definition, function declaration, function definition, and class definitions). Each identifier serves as a unique index within its corresponding database. A unified \texttt{Lookup} function, mentioned in Section \ref{sec:2.3.1}, provides the interface for retrieving specific background knowledge. As discussed in Section \ref{sec:2.3.2}, the \texttt{Need\_To\_Lookup} function is needed to identify the specific type and identifier that require background knowledge. In our experiments, we employ Qwen2.5-72B-Instruct, due to its powerful ability to understand code, to extract protobuf messages, classes, and functions that lack sufficient background information for accurate completion.

For similarity-based RAG, the construction of retrieval databases varies depending on the specific similarity-based retrieval techniques. We utilize BM25S \cite{bm25s}, a recently released efficient BM25 library, to implement the lexical retrieval service. For semantic retrieval, we employ Qdrant\footnote{\url{https://github.com/qdrant/qdrant}}, a high-performance vector database, to construct and manage our retrieval database. The number of retrieved results is set to 4, ensuring that the length of the constructed prompt is less than 2k tokens, which aligns with the context length supported by most LLMs.

\subsubsection{Model Deployment and Inference Settings}

All LLMs and their corresponding tokenizers are obtained from their official Hugging Face repositories and deployed using the vLLM
framework within Docker containers on the WeChat testing platform. To ensure reproducibility and consistency of our experimental results, we maintain consistent inference parameters across all models, setting the temperature to 0 during generation. The hardware configuration for model inference varies according to model size: models with parameters under 200B are tested using 8 NVIDIA A100 GPUs (40GB each), while DeepSeek-Coder-V2-Instruct and DeepSeek-V2.5 are deployed on 8 NVIDIA H20 GPUs (96GB each). For the larger model, DeepSeek-V3, we utilize a cluster of 16 NVIDIA H20 GPUs (96GB each) to accommodate its extensive computational requirements. Except for DeepSeek-V3, which uses FP8, all other LLMs use FP16 precision for inference.
To align with the structure format of our retrieval corpus and benchmark, we design our prompts in Chinese wrapped in C++ comment format for the closed-source code completion.

\section{Empirical Results}
\label{sec-4}
\subsection{RQ1: Effectiveness of RAG}

As shown in Table \ref{tab:tab1}, the experimental results demonstrate that different RAG methods consistently outperform base models across different scales. For instance, the CB/ES metrics of Llama-3.1-8B-Instruct improve from 34.02/46.07 to 39.64/49.35 with function definition retrieval, when using GTE-Qwen-based RAG, Qwen2.5-Coder-14B-Instruct achieves a CB/ES metrics improvement from 29.79/48.56 to 51.12/61.96, representing a 71.60\% and 27.59\% relative increase, respectively. This enhancement pattern can be observed in larger models as well, where DeepSeek-V3 shows an enhancement from 35.23/54.85 to 60.28/73.11 using GTE-Qwen retrieval technique, corresponding to a 71.1\% and 33.3\% increase in performance metrics.

\begin{table*}[htbp]
\centering
\caption{Performance comparison of the combinations between lexical-based and semantics-based retrieval techniques within similarity-based RAG for code completion.}
\label{tab:tab3}
\resizebox{\textwidth}{!}{
\begin{tabular}{c|ccccccc}
\Xhline{1.2pt}
\multirow{2}{*}{\textbf{Model}} & \textbf{BM25} & \textbf{UniXcoder(U)} & \textbf{U+BM25} & \textbf{CoCoSoDa(C)} & \textbf{C+BM25} & \textbf{GTE-Qwen(Q)} & \textbf{Q+BM25} \\
 & CB/ES & CB/ES & CB/ES & CB/ES & CB/ES & CB/ES & CB/ES \\
\hline
\hline
\multicolumn{8}{c}{0.5B+ LLM} \\
\hline
Qwen2.5-Coder-0.5B-Instruct & 27.63/32.45 & 29.98/39.43 & 30.20/35.99 & 30.39/38.68 & 29.86/34.77 & \textbf{34.46}/\textbf{41.67} & 33.84/36.54 \\
\hline
\multicolumn{8}{c}{1B+ LLMs} \\
\hline
OpenCoder-1B-Instruct & \textbf{27.63}/\textbf{32.45} & 21.22/30.79 & 20.36/24.25 & 22.67/29.47 & 15.82/20.09 & 22.12/28.13 & 18.53/20.04 \\
Llama3.2-1B-Instruct & 30.18/32.59 & 25.39/30.02 & 28.18/30.95 & 25.12/30.06 & 28.59/30.56 & 30.93/\textbf{32.64} & \textbf{32.06}/31.50 \\
DS-Coder-1.3B-Instruct & 31.09/39.25 & 25.01/35.54 & 30.12/\textbf{39.42} & 27.76/35.49 & 31.64/38.88 & 34.45/37.06 & \textbf{35.09}/38.04 \\
Qwen2.5-Coder-1.5B-Instruct & 47.78/\textbf{57.62} & 35.82/50.22 & 45.47/53.89 & 41.42/50.85 & 47.04/53.66 & 46.69/56.04 & \textbf{48.78}/55.62 \\
Yi-Coder-1.5B-Chat & \textbf{30.17}/\textbf{38.38} & 23.86/34.07 & 23.40/33.24 & 24.23/34.98 & 23.81/31.88 & 29.06/36.40 & 29.54/34.11 \\
\hline
\multicolumn{8}{c}{3B+ LLMs} \\
\hline
Llama3.2-3B-Instruct & 45.42/47.37 & 38.25/44.80 & 41.43/39.10 & 41.19/45.62 & 46.44/40.33 & \textbf{49.04}/\textbf{51.07} & 48.98/42.94 \\
Qwen2.5-Coder-3B-Instruct & 27.12/44.48 & 17.57/38.52 & 31.84/50.68 & 24.89/42.88 & 35.00/50.94 & 30.63/48.48 & \textbf{35.99}/\textbf{51.36} \\
\hline
\multicolumn{8}{c}{7B+ LLMs} \\
\hline
DS-Coder-7B-Instruct-v1.5 & 43.21/\textbf{55.20} & 37.75/50.37 & 35.06/44.15 & 35.12/47.13 & 26.65/32.10 & \textbf{44.44}/52.36 & 29.63/37.12 \\
Qwen2.5-Coder-7B-Instruct & 45.16/59.36 & 38.36/54.75 & 50.72/63.73 & 44.23/60.96 & 53.58/65.55 & 49.03/62.66 & \textbf{55.57}/\textbf{66.07} \\
Llama3.1-8B-Instruct & 49.80/54.39 & 40.01/49.00 & 49.70/51.98 & 46.22/51.65 & 52.48/53.65 & \textbf{53.47}/\textbf{55.40} & 52.34/52.83 \\
OpenCoder-8B-Instruct & 42.45/32.17 & 34.37/30.08 & 39.80/33.51 & 37.15/30.04 & 39.44/33.09 & 41.38/32.28 & \textbf{42.56}/\textbf{34.21} \\
Yi-Coder-9B-Chat & 51.66/56.14 & 40.28/49.99 & 48.47/52.94 & 42.87/50.49 & 49.63/53.44 & 49.59/55.73 & \textbf{54.05}/\textbf{55.78} \\
\hline
\multicolumn{8}{c}{13B+ LLMs} \\
\hline
CodeLlama-13B-Instruct & 28.00/30.51 & 23.84/28.66 & 30.24/30.65 & 24.63/29.24 & 30.84/30.25 & 26.58/30.70 & \textbf{32.62}/\textbf{32.15} \\
Qwen2.5-Coder-14B-Instruct & 47.28/60.93 & 35.75/52.25 & 52.92/65.00 & 43.01/57.50 & 55.20/66.38 & 49.37/60.80 & \textbf{57.80}/\textbf{66.89} \\
DS-Coder-V2-Lite-Instruct-16B/2.4B & 51.45/57.87 & 40.83/50.75 & 49.50/54.09 & 45.60/55.25 & 50.84/55.36 & \textbf{54.91}/\textbf{57.98} & 54.82/56.43 \\
\hline
\multicolumn{8}{c}{20B+ LLMs} \\
\hline
CodeStral-22B-v0.1 & 47.28/60.93 & 36.16/52.66 & 48.18/58.05 & 43.94/57.05 & \textbf{52.85}/60.24 & 49.37/60.80 & 52.25/\textbf{61.49} \\
Qwen2.5-Coder-32B-Instruct & 55.76/68.89 & 44.16/61.13 & 56.66/69.34 & 49.50/65.54 & 56.90/67.66 & 60.79/71.34 & \textbf{63.73}/\textbf{72.25} \\
DS-Coder-33B-Instruct & 38.91/\textbf{50.19} & 34.02/47.96 & 34.28/45.21 & 34.50/46.23 & 28.45/35.45 & \textbf{39.40}/47.43 & 33.84/41.83 \\
CodeLlama-34B-Instruct & 27.53/36.48 & 18.01/31.58 & 21.43/31.13 & 22.12/33.93 & 22.09/31.72 & \textbf{28.35}/\textbf{37.42} & 23.61/32.53 \\
\hline
\multicolumn{8}{c}{70B+ LLMs} \\
\hline
CodeLlama-70B-Instruct & \textbf{19.17}/\textbf{26.95} & 15.23/21.34 & 14.27/17.76 & 16.05/20.30 & 15.59/19.87 & 16.26/20.70 & 13.72/18.10 \\
Llama-3.3-70B & 50.21/62.37 & 39.33/55.02 & 49.84/61.57 & 45.53/58.86 & 50.42/59.80 & 52.64/\textbf{64.60} & \textbf{54.00}/64.37 \\
Qwen2.5-72B & 50.21/62.45 & 45.57/60.38 & 52.99/62.67 & 48.16/61.96 & 53.55/61.27 & 56.05/\textbf{65.35} & \textbf{59.03}/64.89 \\
\hline
\multicolumn{8}{c}{200B+ LLMs} \\
\hline
DS-Coder-V2-Instruct-236B/21B & 38.35/57.95 & 44.56/61.60 & 56.59/68.48 & 48.58/63.33 & 56.34/68.16 & 48.58/63.33 & \textbf{62.25}/\textbf{71.36} \\
DS-V2.5-236B/21B & 55.67/69.18 & 44.70/61.69 & 55.45/68.10 & 48.16/63.85 & 56.68/68.20 & 55.29/68.21 & \textbf{61.40}/\textbf{71.12}  \\
DS-V3-671B/37B & 55.14/68.55 & 44.75/62.27 & 57.48/70.97 & 50.43/65.40 & 60.48/72.34 & 60.28/73.11 & \textbf{63.62}/\textbf{75.26}  \\
\Xhline{1.2pt}
\end{tabular}
}
\end{table*}

Among the identifier-based RAG methods, function definition retrieval consistently yields the highest performance gains. This is particularly evident in models like Qwen2.5-Coder-32B-Instruct, where function definition retrieval improves the base CB/ES metrics from 38.05/57.89 to 42.23/60.44, outperforming other background knowledge, such as message definition and class definition. In the similarity-based RAG, BM25 and GTE-Qwen-based retrieval techniques demonstrate superior performance, with BM25 achieving CB/ES metrics of 55.67/69.18 and GTE-Qwen reaching 55.29/68.21 for DeepSeek-V2.5, compared to other similarity-based retrieval techniques like CodeBERT and UniXcoder.

The comparative analysis between the two types of RAG methods reveals a clear advantage for similarity-based RAG. The superiority is consistently observed across different models and scales. For example, Qwen2.5-Coder-1.5B-Instruct achieves a maximum CB/ES metrics of 37.28/50.77 within identifier-based RAG, while similarity-based RAG pushes the performance to 46.69/56.04. DeepSeek-V3 reaches 42.24/61.75 performance with identifier-based RAG but achieves 60.28/73.11 with similarity-based RAG, representing a substantial improvement of 42.7\% and 18.4\% respectively.

\begin{tcolorbox}
[width=\linewidth-2pt,boxrule=3pt,top=3pt, bottom=1pt, left=3pt,right=3pt, colback=gray!20,colframe=gray!25]
\textbf{Finding 1:}  
Both types of RAG methods can consistently improve code completion performance across different models and scales in closed-source repositories. 
Moreover, compared to identifier-based RAG, similarity-based RAG substantially performs better in enhancing code completion quality.
\end{tcolorbox}


\subsection{RQ2: Impact of Retrieval Techniques}
The experimental results in Table \ref{tab:tab2} reveal that among semantic retrieval techniques, CodeBERT consistently underperforms compared to UniXcoder, CoCoSoDa, and GTE-Qwen. This performance gap may be attributed to the differences in pre-training objectives: while CodeBERT relies solely on MLM and RTP, the other three models are specifically optimized for retrieval tasks through contrastive learning, which better captures code semantics and similarity relationships.
An example is that Qwen2.5-Coder-14B-Instruct achieves CB/ES metrics of 51.12/61.96 with incomplete queries, still surpassing 23.23/29.54 gained by CodeBERT.
Different from semantic retrieval techniques that require training, BM25, as a lexical retrieval technique, exhibits remarkable effectiveness through simple term-matching mechanisms. Our experimental results show that BM25-based RAG consistently achieves strong performance across various model scales and architectures.

\begin{figure*}[h]
  \centering
  \includegraphics[width=\textwidth]{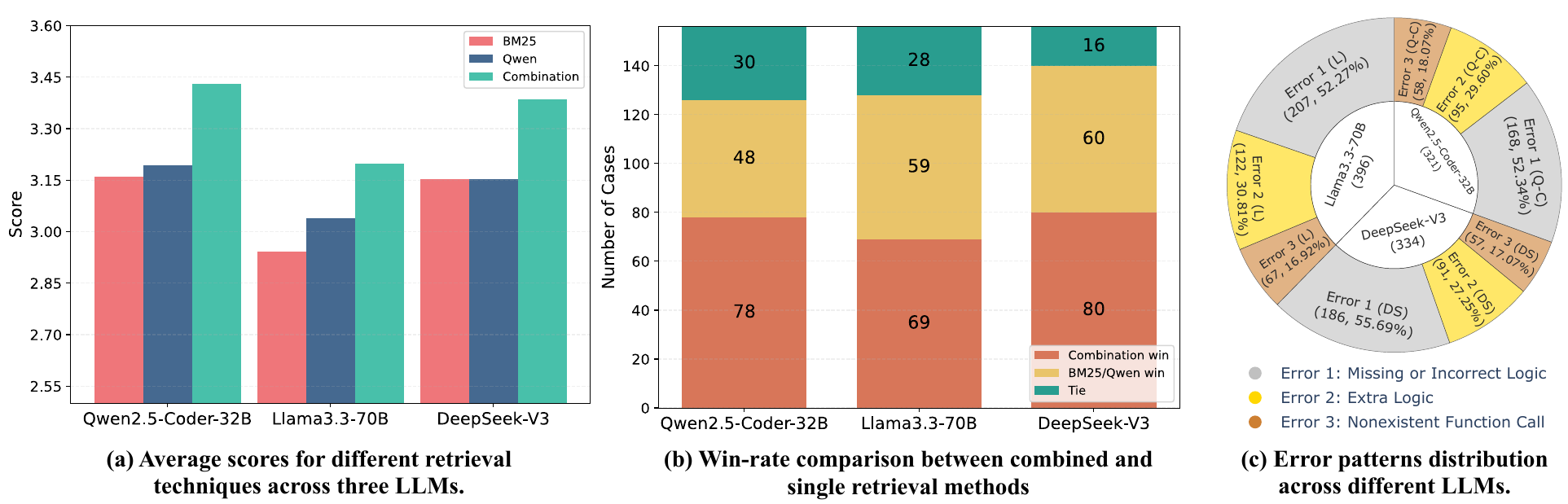}
  \caption{The results analysis of developer survey.}
  \label{fig:user}
\end{figure*}

In code completion tasks, only incomplete code context can be used as queries during retrieval, which creates a misalignment with the representation learning process of retrieval techniques.
As shown in Table \ref{tab:tab2}, UniXcoder and CoCoSoDa demonstrate superior performance with the entire code snippets as queries, suggesting their potential ability to further improve RAG performance on code completion. BM25 also shows consistent improvements when using complete queries across all LLMs, with CB/ES metrics increasing from 31.43/41.25 to 36.40/43.48 in the 0.5B scale.
In contrast, GTE-Qwen consistently demonstrates superior performance across different LLMs with incomplete code contexts as queries, particularly evident in larger LLMs. For example, when applied to DeepSeek-V3, GTE-Qwen achieves the highest CB/ES scores of 60.28/73.11, surpassing other retrieval techniques. The powerful retrieval ability may be attributed to its architecture, larger scale, and strong code-specific pre-training, making it particularly well-suited for code completion tasks.

\begin{tcolorbox}
[width=\linewidth-2pt,boxrule=3pt,top=3pt, bottom=1pt, left=3pt,right=3pt, colback=gray!20,colframe=gray!25]
\textbf{Finding 2:} Lexical retrieval technique consistently exhibits remarkable performance across different query formulations.
While most retrieval techniques can be further improved by complete queries, GTE-Qwen demonstrates better performance with incomplete queries, making it particularly suitable for the code completion task.
\end{tcolorbox}

\subsection{RQ3: Exploration on Retreived Results}

Due to the superior performance of similarity-based RAG, we conduct an exploratory analysis of retrieved results from lexical and semantic retrieval techniques. The results reveal minimal overlap between BM25 and semantic techniques: out of 100 test examples in our benchmark, there are 76, 74, and 64 completely distinct retrieved samples when comparing BM25 with UniXcoder, CoCoSoDa, and GTE-Qwen, respectively. This difference in retrieval distributions motivates us to explore the combination of these techniques. 

As shown in Table \ref{tab:tab3}, the effectiveness of combining BM25 with different semantic retrieval techniques becomes more pronounced as model size increases. In the 200B+ scale, the BM25+GTE-Qwen combination achieves CB/ES metrics of 63.62/75.26 for DeepSeek-V3, substantially outperforming both individual techniques. Similarly, other semantic retrieval techniques also benefit from the combination of BM25. DeepSeek-V2.5 improves CB/ES metrics from 48.16/63.85 to 56.68/68.20 when combined with BM25. Notably, the advantage of BM25+GTE-Qwen combination is particularly striking for Qwen2.5-32B, which achieves impressive CB/ES scores of 63.73/72.25, rivaling or even surpassing models with significantly larger parameters such as DeepSeek-V2.5 and DeepSeek-V3. However, for smaller models (below 7B), the combination shows limited or even negative impact, suggesting that the complementary benefits of hybrid retrieval methods are more effectively leveraged by larger models.

\begin{tcolorbox}
[width=\linewidth-2pt,boxrule=3pt,top=3pt, bottom=1pt, left=3pt,right=3pt, colback=gray!20,colframe=gray!25]
\textbf{Finding 3:} Lexical and semantic retrieval techniques exhibit distinct retrieval results distribution and demonstrate complementary characteristics in larger-scale models (7B+). With their combination, especially BM25+GTE-Qwen, similarity-based RAG achieves optimal performance in most LLMs.
\end{tcolorbox}

\section{Discussion}
\label{sec-5}
\subsection{Developer Survey}
To further validate the superior outcomes achieved by integrating lexical and semantic retrieval results within similarity-based RAG, we conduct a developer survey involving three developers from our group (excluding the authors). This study aims to assess the quality of code completions generated using various retrieval techniques in similarity-based RAG, including BM25, GTE-Qwen, and a combination of both. The evaluation is performed on a random selection of 52 examples and three LLMs from the Qwen, Llama, and DeepSeek families, which demonstrated the best performance with the combined use of BM25 and GTE-Qwen.

Specifically, we design an evaluation website for the study. The current code context and the ground truth code are provided as reference. The participating developers could assess the quality of the generated code by comparing the generated completion with the code context and ground truth. 
The scoring system uses a 1-5 scale, where 1 indicates significant differences from the original code with serious errors, and 5 represents near-perfect alignment with the original code and excellent quality. Additionally, based on our preliminary analysis of code completion failures, we identified three common issue categories that we provide as predefined options in the evaluation interface: \textbf{code logic errors}, \textbf{external code dependencies missing}, and \textbf{reference to non-existent functions}. These additional evaluation criteria help us better understand the root causes of low-quality code completion.

The developer survey reveals several important insights about different retrieval techniques and their effectiveness. The combination of BM25 and GTE-Qwen consistently achieves higher scores than using either technique alone across all three LLMs. This advantage is particularly evident in the win-rate analysis, where the combined technique outperforms single technique in about half of all test cases. While DeepSeek-V3 shows no strong preference between the retrieved results of BM25 and GTE-Qwen individually, both Qwen2.5-32B-Instruct and Llama3.3-70B-Instruct perform better with GTE-Qwen-based RAG.
The error analysis reveals a consistent pattern across all LLMs. Missing or Incorrect Logic dominates the error types at around 52\%. This suggests that improving logical reasoning capabilities should be a priority for future development. Llama3.3-70B-Instruct exhibits slightly more errors overall compared to other LLMs, but the general distribution of error types remains similar across all three LLMs.

\subsection{Implication of Findings}

Our findings have several important implications for the development and application of RAG for code completion in closed-source scenarios:

\textbf{Leveraging open-source LLMs with RAG for proprietary code development:} 
Our results demonstrate that RAG can effectively leverage knowledge from closed-source codebases to improve the code completion performance of open-source LLMs. It is particularly valuable for proprietary development environments where access to extensive training data may be limited.
The transparent and accessible nature of open-source models addresses privacy concerns in the development process, making them particularly suitable for real-world industrial applications.
Moreover, the similarity-based RAG proves to be a more effective solution for improving the performance of various LLMs compared to the identifier-based RAG.

\textbf{Exploring task-specific retrieval techniques:}
We identify a gap between the training and application scenarios of semantic retrieval techniques. These models are typically trained on complete code snippets. However, code completion tasks require retrieving complete code snippets based on incomplete code fragments. This misalignment suggests two potential directions for improvement: (1) developing specialized retrieval techniques optimized for incomplete queries by reducing the semantic distance between partial and complete code snippets, and (2) utilizing more powerful retrieval models like GTE-Qwen that naturally accommodate incomplete query scenarios without additional fine-tuning.

\textbf{Combining different types of retrieval techniques within RAG:}
Previous works often employ one type of retrieval technique for initial candidate selection, followed by another type of retrieval technique for re-ranking \cite{DBLP:conf/naacl/GlassRCNCG22,dong2024don,finardi2024chronicles}. Our findings suggest a more nuanced relationship between the two types of retrieval techniques. The distinct yet complementary nature of lexical and semantic retrieval techniques indicates potential for further performance improvements in similarity-based RAG systems through their strategic combination, beyond the traditional retrieve-then-rerank pipeline.

\subsection{Threats to Validity}

\textbf{Internal Validity.} In our exploration of RAG using various LLMs and retrieval techniques, the performance of these deep learning-based models can be influenced by multiple factors, including parameter settings and hardware devices. To address this potential threat, we maintain consistency by using default parameter configurations across all models and set the temperature parameter to 0 during LLMs inference to ensure reproducible results.

\textbf{External Validity.} Our experiments are conducted on the specific enterprise codebase in WeChat group, which might exhibit distinct characteristics from other organizations' codebases. To mitigate this limitation, we select a diverse set of projects (total 1,669) covering different domains and development periods within our codebase. This dataset encompasses various development practices, coding standards, and business scenarios, providing a comprehensive representation of software development patterns in a closed-source environment.





\textbf{Construct Validity.}
We use automated metrics (CodeBLEU and Edit Similarity) to measure code quality, but since code completion tools are ultimately used by developers, these metrics might not fully capture the semantic correctness and functionality of generated code in real development scenarios. To address this limitation, we supplement human evaluation with a developer survey, categorizing potential error types and identifying future research directions for optimization.




\section{Related Wrok}
\label{sec-6}
Code completion, which aims to predict subsequent code elements based on the existing context, is a crucial task for improving developer productivity in software engineering. Early approaches primarily rely on statistical methods to implement code completion functionality \cite{statistical-cc}. Recently, with the advancement of deep learning, particularly LLMs, code completion has achieved superior performance in production environments and provide developers meaningful suggestions \cite{sun2024neural,zhu2024exploring}.
Several recent works have focused on improving code completion through various context selection techniques. Liang et al. \cite{repofuse} extract dependency definitions from the current context and retrieve similar code snippets from a code repository, aggregating both to construct prompts that help LLMs better understand the context for code completion. Cheng et al. \cite{24-Cheng} introduce code dependencies through data flow graphs in a directed manner. Liu et al. \cite{liu2024graphcoder} locate context segments relevant to code completion using structured patterns and implement a reranking algorithm based on decay-with-distance sub-graph edit distance. Additionally, Liu et al. \cite{liu2024stall+} incorporate multiple static analysis methods across different stages of code completion to enhance the reliability of completed code.


\section{Conclusion}
\label{sec-7}
In this paper, we conduct a systematic investigation of retrieval-augmented generation (RAG) for code completion in closed-source repositories. Through comprehensive experiments on 26 open-source LLMs ranging from 0.5B to 671B parameters, we demonstrate the consistent effectiveness of both identifier-based and similarity-based RAG methods. Our in-depth analysis of similarity-based RAG reveals that BM25 and GTE-Qwen achieve superior performance in code completion. Furthermore, we explore the relationship between lexical and semantic retrieval techniques, identifying the BM25+GTE-Qwen combination as the optimal improvement strategy. We summarize our findings and provide valuable insights for researchers and practitioners to apply RAG methods for code completion systems in their proprietary environments.

\normalem
\bibliographystyle{IEEEtran}
\bibliography{reference}

\end{document}